\documentclass[
]{ceurart}

\usepackage{listings}
\usepackage{soul}
\usepackage{tcolorbox}
\begin{document}

\copyrightyear{2024}
\copyrightclause{Copyright for this paper by its authors.
  Use permitted under Creative Commons License Attribution 4.0
  International (CC BY 4.0).}

\conference{LLM4Eval: The First Workshop on Large Language Models for Evaluation in Information Retrieval, 18 July 2024, Washington DC, United States.}

\title{Exploring Large Language Models\\
for Relevance Judgments in Tetun}

%\tnotetext[1]{You can use this document as the template for preparing your
 % publication. We recommend using the latest version of the ceurart style.}

\author[1,2]{Gabriel de Jesus}[%
orcid=0000-0003-4392-2382,
email=gabriel.jesus@inesctec.pt,
]
\cormark[1]
\address[1]{INESC TEC - Institute for Systems and Computer Engineering, Technology and Science, Portugal}
\address[2]{FEUP - Faculty of Engineering, University of Porto, Portugal}

\author[1,2]{Sérgio Nunes}[%
orcid=0000-0002-2693-988X,
email=sergio.nunes@fe.up.pt,
url=https://web.fe.up.pt/~ssn/,
]

\cortext[1]{Corresponding author.}

\begin{abstract}
  The Cranfield paradigm has served as a foundational approach for developing test collections, with relevance judgments typically conducted by human assessors. However, the emergence of large language models (LLMs) has introduced new possibilities for automating these tasks. This paper explores the feasibility of using LLMs to automate relevance assessments, particularly within the context of low-resource languages. In our study, LLMs are employed to automate relevance judgment tasks, by providing a series of query-document pairs in Tetun as the input text. The models are tasked with assigning relevance scores to each pair, where these scores are then compared to those from human annotators to evaluate the inter-annotator agreement levels. Our investigation reveals results that align closely with those reported in studies of high-resource languages.
\end{abstract}


\maketitle

%%% INTRODUCTION %%%
\section{Introduction} \label{introduction}

The advancement of information retrieval (IR) systems depends on the availability of reliable test collections to assess their effectiveness. The traditional approach for developing these collections follows the Cranfield paradigm \cite{cleverdon-1976}, which became widely recognized through the Text REtrieval Conference (TREC) series of large-scale evaluation campaigns \cite{harman-et-al-1992}. In TREC guidelines, a test collection comprises a document collection, a set of topics, and corresponding relevance assessments. The relevance judgment tasks are typically carried out by human assessors, a process that is both time-consuming and costly.
 
To tackle the aforementioned problems, the IR community has been investigating the feasibility of automatically generated relevance judgments for developing test collections. With the advent of large language models (LLMs), which have demonstrated proficiency in various tasks, new possibilities for conducting automated relevance judgments have emerged, demonstrating ongoing improvement in the quality of automated relevance judgment tasks as LLMs continue to evolve. 

Studies have consistently shown that LLMs are effective in automated relevance assessment tasks, providing their cost-effectiveness solutions with judgment agreement comparable to human assessors. Faggioli et al.~\cite{fraggioli-et-al-2023} argued that although further improvement in LLMs capabilities is necessary for fully automated relevance judgments, LLMs are already capable of assisting humans in this task. Additionally, a recent study by Bueno et al.~\cite{bueno-et-al-2024} reported a consistent improvement in automated relevance judgments with an average Cohen's Kappa score of 0.31 for annotation agreement between humans and LLMs, which are inline with the findings of Faggioli et al.~\cite{fraggioli-et-al-2023}. However, these studies primarily focus on high-resource languages, such as English and Brazilian Portuguese, leaving the applicability of LLMs in low-resource language (LRL) contexts as an open question.     

In this study, we explore the use of LLMs to automate relevance judgment tasks in Tetun, a LRL spoken by over 923,000 people in Timor-Leste~\cite{jesus-2023}. We used an existing test collection comprising 6,100 relevance judgments constructed utilizing documents from the Labadain-30k+ dataset \cite{labadain30k-dataset-2024}. The relevance judgments for this collection were conducted by native Tetun speakers. These query-document pairs were provided to the LLMs to assign relevance scores for each. We compared these scores with those from human annotations and observed inter-annotator agreement levels. The results revealed an inter-annotator agreement of Cohen's kappa score of 0.2634 when evaluated using the 70B variant of the LLaMA3 model \cite{touvron-et-al-llama-2023}. This finding demonstrates the feasibility of using LLMs in LRL scenarios to automate the relevance judgment tasks.

The remaining sections of this paper are organized as follows. Section~\ref{sec:related-works} describes related work. An overview of the collection used in this study is outlined in Section~\ref{sec:collection-overview}. Then, Section~\ref{sec:relevance-judgments} details the experiment of using LLMs for automating relevance judgments. Section~\ref{sec:results-discussions} presents the results obtained and their discussion. Finally, Section~\ref{sec:conclusions-future-work} summarizes our conclusion and possible future work.

%%% RELATED WORKS %%%
\section{Related Work}\label{sec:related-works}

Test collections are the most important component used for evaluating the effectiveness of IR systems. For high-resource languages, these collections are typically made available through large-scale campaigns such as Text REtrieval Conference (TREC)\footnote{\url{https://trec.nist.gov}}, the Conference and Labs of the Evaluation Forum (CLEF)\footnote{\url{https://www.clef-initiative.eu}}, the NII Testbeds and Community for Information Access Research project (NTCIR)\footnote{\url{http://research.nii.ac.jp/ntcir/index-en.html}}, and the Forum for Information Retrieval Evaluation (FIRE)\footnote{\url{http://fire.irsi.res.in/}}.  

The TREC-style approach, derived from the Cranfield paradigm, is commonly adopted for developing test collections, including for low-resource languages (LRLs), where human assessors conduct the relevance judgment tasks \cite{sahu-pal-2023, chavula-suleman-2021, esmaili-et-al-2013, aleahmad-et-al-2009}. However, the fast pace of research and innovation, particularly with the emergence of LLMs, has significantly transformed natural language processing (NLP). Within the IR domain, studies have demonstrated that automated relevance judgments using LLMs can yield results comparable to traditional methods, and these outcomes have consistently improved as LLMs have evolved. Initially, Faggioli et al.~\cite{fraggioli-et-al-2023} explored the potential application of LLMs to fully-automated relevance judgment tasks. They analyzed the judgment results from the TREC 2021 Deep Learning track \cite{craswell-et-al-trec-dl-task-2021} and compared them with LLM-based relevance assessments generated using GPT-3.5 of OpenAI\footnote{\url{https://openai.com}}. Their findings revealed a Cohen's kappa score of 0.26 for inter-annotator agreement between human and LLM, indicating a fair level of agreement. Thus, they argued that LLMs are already capable of assisting humans in relevance judgment tasks, despite further improvement in LLM capabilities are necessary for fully automated relevance judgments.

Later, Thomas et al.~\cite{thomas-et-al-2023} reported that LLMs demonstrated accuracy comparable to human labelers when deployed for large-scale relevance labeling at Bing. Their work utilized the GPT-4 model \cite{gpt4-report} and incorporated data from the TREC Robust04 track \cite{voorhees-trec-robust-2004}, showing that LLMs achieved a Cohen's kappa score ranging from 0.20 to 0.64 for agreement between humans and LLM across various tasks. In a recent study, Bueno et al.~\cite{bueno-et-al-2024} in their study while constructing a test collection for Brazilian Portuguese, reported consistent improvement and findings comparable to those of Thomas et al.~\cite{thomas-et-al-2023} and Faggioli et al.~\cite{fraggioli-et-al-2023}, with automated relevance judgments yielding an average Cohen's Kappa score of 0.31 for annotation agreement between humans and LLMs. 

Despite these advancements, uncertainties persist about the feasibility of using LLMs to automatically generating relevance judgments for LRLs. Thus, our research focuses on exploring this potential application in LRL scenarios, specifically in Tetun.

%%% COLLECTION OVERVIEW %%%
\section{Collection Overview}\label{sec:collection-overview}

In this experiment, we utilized the existing Tetun test collection\footnote{This collection has not yet been published.} developed according to TREC guidelines. The following subsections detail the test collection used in this work.

\subsection{Documents}

Documents of the Tetun test collection are derived from the Labadain-30k+ dataset, which consists of 33,550 documents in Tetun \cite{labadain30k-dataset-2024}. This dataset was acquired from the web and encompassed a broad array of categories, including news articles, Wikipedia entries, legal and government documents, research papers, and more \cite{de-jesus-nunes-2024-labadain}. A summary of the document collection is provided in \autoref{tab:docs-summary}.

\begin{table}
  \caption{Summary of Document Collection. *Tokens Comprise Words and Numbers, Excluding Punctuation and Special Characters.}
  \label{tab:docs-summary}
  \begin{tabular}{lr}
    \toprule
    Description & Value\\
    \midrule
    Total number of documents & 33,550\\
    Size of collection & 84MB\\
    Total number of tokens* & 12,300,237 \\ 
    Total number of distinct tokens & 162,466\\
  \bottomrule
\end{tabular}
\end{table}

\subsection{Queries}\label{subsec-queries}

The collection consists of 61 queries developed by five volunteer students, all Timoreses and native Tetun speakers. The queries are originated from the logs of Timor News\footnote{\url{https://www.timornews.tl}}, an online newspaper based in Dili, Timor-Leste. Statistics about the queries are presented in \autoref{tab:queries-summary}.

\begin{table}
  \caption{Summary of Queries.}
  \label{tab:queries-summary}
  \begin{tabular}{lr}
    \toprule
    Description & Value\\
    \midrule
    Total number of queries & 61\\
    Minimum number of words per query & 3\\
    Maximum number of words per query & 5\\
    Average numbers of words per query & 3.42\\
  \bottomrule
\end{tabular}
\end{table}

\subsection{Relevance Judgments}

Relevance judgments were conducted by the same five Timorese students. These students were tasked with evaluating the relevance of query-document pairs. The pairs were classified into four graded levels of topical relevance: irrelevant, marginally relevant, relevant, and highly relevant, as proposed by Sormunen \cite{sormunen-2002}. The inter-annotator agreement achieved an average a Cohen's kappa score of 0.4236 and the details of the resulting test collection are presented in \autoref{tab:detail-human-jugment}.

\begin{table}
  \caption{Human Relevance Judgment Results.}
  \label{tab:detail-human-jugment}
  \begin{tabular}{lrr}
    \toprule
    Relevance Level & Total & Proportion\\
    \midrule
    3 - Highly relevant & 710 & 11.64\%\\
    2 - Relevant & 1,102 & 18.07\%\\
    1 - Marginally relevant & 476 & 7.80\%\\
    0 - Irrelevant & 3,812 & 62.49\%\\
  \bottomrule
\end{tabular}
\end{table}

%%% AUTOMATED RELEVANCE JUDGMENTS %%%
\section{Relevance Judgments Using LLMs} \label{sec:relevance-judgments}

\subsection{Overview}

Several studies have already utilized the GPT-3.5 and GPT-4 models from OpenAI for automating relevance judgment tasks \cite{fraggioli-et-al-2023, thomas-et-al-2023, bueno-et-al-2024}. However, due to the costs associated with these LLMs, our study explores an alternative by employing the freely available 70B variant of LLaMA3, released by Meta on April 18, 2024 \cite{metallama3}. We conduct automated relevance judgments using the Tetun test collection detailed in \autoref{sec:collection-overview}, to compare their inter-annotator agreement levels.

Additionally, to evaluate whether the free LLaMA3 model of 70B variant can outperform certain paid LLMs in relevance assessment tasks, specifically within the Tetun context, we have selected two paid models for comparison: the Haiku variant of Claude 3 from Anthropic\footnote{\url{https://www.anthropic.com}}, and the Turbo variant of GPT-3.5 from OpenAI. A summary of the models used, along with their associated costs, is presented in \autoref{tab:cost-llms}.

\begin{table}
  \caption{Price Associated with LLMs.}
  \label{tab:cost-llms}
  \begin{tabular}{lrr}
    \toprule
    Model & Input  & Output\\
    \midrule
    LLaMA 3 & --- & ---  \\
    Claude 3 Haiku & \$0.25 / 1M tokens & \$1.25 / 1M tokens\\
    GPT-3.5-turbo-0125 & \$0.50 / 1M tokens & \$1.50 / 1M tokens\\
  \bottomrule
\end{tabular}
\end{table}

To assess the suitability of the chosen LLMs for Tetun, including the two paid models, we conducted preliminary tests that involved translating Tetun text into English. This step was essential given that the query-document pairs are written in Tetun. Examples of these translated outputs are presented in \autoref{tab:llms-translate-tetun}, showing that LLaMA3 inaccurately translated two words, as indicated by strike-through markings.

To evaluate the quality of the translated text generated by the LLMs, we randomly selected a sample of five documents from the query-document pairs (see example in \autoref{tab:query-doc-pairs}), and translated them into English. These human translations served as reference points for evaluation. The assessment using the BLEU metric \cite{papineni-etal-2002-bleu}, demonstrates that both paid models outperformed LLaMA3 in translating Tetun to English, as shown in \autoref{tab:bleu-score}.

\begin{table}
  \caption{Examples of Tetun to English Translations Provided by LLMs.}
  \label{tab:llms-translate-tetun}
  \begin{tabular}{p{0.30\columnwidth}p{0.60\columnwidth}}
  Tetun text: & \textbf{Juventude hetan virus infeksaun HIV/SIDA tanba menus informasaun.}\\
    \toprule
    Model & Translation\\
    \midrule
    LLaMA 3 & Youth \st{against} virus infection of HIV/AIDS \st{and} lack of information.\\
    Claude 3 Haiku & Youth get HIV/AIDS virus infection due to lack of information.\\
    GPT-3.5-turbo-0125 & The youth are getting infected with HIV/AIDS because of lack of information.\\
    \bottomrule
  \end{tabular}
\end{table}

\begin{table}
  \caption{Evaluation of Translation Quality Produced by LLMs Using BLEU.}
  \label{tab:bleu-score}
  \begin{tabular}{lr}
    \toprule
    Model & BLEU\\
    \midrule
    LLaMA 3 & 0.3849\\
    Claude 3 Haiku & 0.4167 \\
    GPT-3.5 Turbo & 0.6525 \\
  \bottomrule
\end{tabular}
\end{table}

However, given that relevance judgment tasks require not only direct translation but also a nuanced level of understanding, we compared the selected models' multi-task language understanding capabilities using the Massive Multitask Language Understanding (MMLU)~\cite{hendrycks-et-al-2021} based on the MMLU benchmark leaderboard \cite{mmlu-leaderboard}. A summary of these LLMs' performance on MMLU is outlined in \autoref{tab:llms-mmlu}. It shows that in the few-shot scenario with five examples, LLaMA 3 surpassed Claude 3 Haiku by an average of +5 percentage points and GPT-3.5 Turbo by +10.2 percentage points.

\begin{table}
  \caption{LLM Capabilities of Multi-task Language Understanding on MMLU.}
  \label{tab:llms-mmlu}
  \begin{tabular}{lr}
    \toprule
    Model & Average (\%)\\
    \midrule
    LLaMA 3 (5-shot) & 80.2  \\
    Claude 3 Haiku (5-shot) & 75.2 \\
    GPT-3.5 Turbo (5-shot) & 70.0 \\
  \bottomrule
\end{tabular}
\end{table}

\subsection{Experiment with Tetun}

To automate relevance judgments using LLMs, we utilized few-shot prompting, adopting a structure similar to that employed by Bueno et al. \cite{bueno-et-al-2024}. Our prompt, along with an example, is illustrated in Prompt \ref{promptbox}, and the full prompt is outlined in \autoref{annex:prompt}. We provided the LLMs with a total of 6,100 query-document pairs and tasked the LLMs with assigning a relevance score to each. Examples of these query-document pairs are depicted in \autoref{tab:query-doc-pairs}.

\begin{table}
  \caption{Examples of Query-Document Pairs.}
  \label{tab:query-doc-pairs}
  \begin{tabular}{p{0.30\columnwidth}p{0.60\columnwidth}}
    \toprule
    Query & Document \\
    \midrule
    Prevensaun moras HIV/SIDA & UNFPA Sei Koopera ho MS Hodi halo Prevensaun ba Moras HIV/SIDA\\
    Prevensaun moras HIV/SIDA & KNK-HIV/SIDA Sensibiliza Informasaun HIV/SIDA Ba Traballador KSTL\\
    Prevensaun moras HIV/SIDA & Autoridade Lokál Partisipa Workshop Prevensaun Moras HIV/SIDA\\
  \bottomrule
\end{tabular}
\end{table}

Given that the existing Tetun test collection employs four-level relevance scores ranging from 0 to 3, we provided the LLMs with query-document pairs alongside four examples, one for each relevance score. These examples used the same queries as those utilized in the pilot testing phase by human assessors, including the relevance score and the reasoning behind each score. For each request, we asked the LLMs to assign one of the four scores and provide the reasoning for their assigned score.

For the 70B variant of the LLaMA 3 model, which requires a substantial amount of memory to run locally, specifically a minimum of 40 GB of RAM as indicated on Ollama\footnote{\url{https://ollama.com/library/llama3:70b}}, we utilized the free API version of the cloud infrastructure provided by Groq\footnote{\url{https://console.groq.com/settings/billing}} to execute this model. However, the scripts for automated relevance judgments for all models were executed locally.

\newtcolorbox[list inside=promptbox,auto counter,number within=section]{PromptBox}{colback=red!5!white, colframe=red!50!black,title={Prompt \thetcbcounter: Example of the System Prompt.},label=promptbox}

\begin{PromptBox}

You are an expert assessor and you are tasked with assessing the relevance between the input query and its corresponding document, assigning a score from 0 to 3. A score of 0 indicates irrelevant; 1, marginally relevant; 2, relevant; and 3, highly relevant.\\

Example:\\\newline
\textbf{query:} ``Kursu mestradu no pós-graduasaun UNTL''\\
\textbf{document:} ``Kursu Desportu UNTL sei realiza graduasaun dahuluk tinan ne’e''\\
\textbf{reason:} ``The query is about postgraduate and master's courses at UNTL, whereas the document focuses on a sports course. Despite both courses in the query and document being offered at UNTL, the sports course in the document is not specifically designed for postgraduate or master's levels. Thus, the document is only marginally relevant.''\\ 
\textbf{score:} 1\\

The query and document to be evaluated are the following:\\\newline
\textbf{query:} $\{query\}$\\
\textbf{document:} $\{document\}$\\\newline
Your response must be in JSON format with the first field is ``reason'', explaining your reasoning, and the second field is ``score''.
\end{PromptBox}

We initiated the experiment with the LLaMA3 70B model, as it was our primary target for comparing annotator agreement level with human annotators. We tested this model using temperatures of 0.0 and 0.5, respectively. The concept of comparing different model temperatures in inter-annotator agreement was inspired by the work of Ma et al. \cite{ma-et-al-2024}, who applied LLMs for relevance judgments in Chinese legal case retrieval. When we increased the temperature of LLaMA3 70B model, the results were not satisfactory. Therefore, we opted to use a zero temperature setting in the other paid models for comparison.

%%% RESULTS AND DISCUSSIONS %%%
\section{Results and Discussions}\label{sec:results-discussions}

In the experiment with the LLaMA3 70B model set at zero temperature, we obtained an inter-annotator agreement of Cohen's kappa score of 0.2634 with human annotators. After increasing the temperature to 0.5, the inter-annotator agreement slightly decreased to 0.2594 (a reduction of -0.004). This finding aligns with the research by Ma et al. \cite{ma-et-al-2024}, where their Cohen's kappa score of inter-annotator agreement levels between humans and LLMs also marginally decreased when they raised the temperature from 0.4 to 0.7 in evaluations of material facts.

Consequently, we opted for a zero temperature setting when conducting relevance judgments with the Claude3 Haiku and GPT-3.5 Turbo models. Comparisons of the inter-annotator agreement levels between LLMs and human annotators are presented in \autoref{tab:cohen-k-annotators}. These results show that the LLaMA3 70B model achieved a highest Cohen's kappa score, indicating the most substantial agreement with human annotators compared to both paid models. Among the paid models, GPT-3.5 Turbo exhibited a slightly higher Cohen's kappa score than Claude3 Haiku (a \textit{k} score increase of +0.0012). Thus, despite the superior performance of the paid models in translating Tetun into English, this finding suggests that a deeper level of language understanding is more crucial in automated relevance judgment tasks.

\begin{table}
  \caption{Cohen's Kappa Correlations Between Human Assessors and LLMs.}
  \label{tab:cohen-k-annotators}
  \begin{tabular}{lrrr}
    \toprule
    & LLaMA3 70B  & Claude3 Haiku & GPT-3.5-turbo-0125\\
    \midrule
    Human & \textbf{0.2634} & 0.2450 &  0.2462 \\     
  \bottomrule
\end{tabular}
\end{table}

As a result, our finding using LLaMA3 70B model is closely aligned with the initial results reported by Faggioli et al.~\cite{fraggioli-et-al-2023}, and are consistent with the findings of Bueno et al.~\cite{bueno-et-al-2024} and Thomas et al.~\cite{thomas-et-al-2023}. Comparisons of these findings regarding the use of LLMs to automate relevance judgments are presented in \autoref{tab:comparison-llms-sota}.

\begin{table}
  \caption{Comparison of the Inter-Annotator Agreement Levels in the Relevance Judgment Tasks using LLMs.}
  \label{tab:comparison-llms-sota}
  \begin{tabular}{llr}
    \toprule
     & LLM & Cohen's \textit{k} Score\\
    \midrule
    Bueno et al.~\cite{bueno-et-al-2024} & GPT-4 & 0.31 \\
    Thomas et al.~\cite{thomas-et-al-2023} & GPT-4 & 0.26 --- 0.64 \\
    Faggioli et al.~\cite{fraggioli-et-al-2023} & GPT 3.5 & 0.26\\
    \textbf{Ours} & LLaMA3 70B & 0.26 \\
  \bottomrule
\end{tabular}
\end{table}

Furthermore, our experiments took an average of approximately 3.56 hours to complete the relevance judgment tasks for each model. The costs associated with the two paid models are detailed in \autoref{tab:expense-llm}. Given that GPT-3.5 Turbo is priced \$0.25 higher per use than Claude 3 Haiku for every 1 million input and output tokens, the expenses for GPT-3.5 were higher than those for Claude 3 Haiku.

\begin{table}
  \caption{Expense on LLMs for Relevance Judgment Tasks.}
  \label{tab:expense-llm}
  \begin{tabular}{lr}
    \toprule
    Model & Expense\\
    \midrule
    LLaMA 3 & ---\\
    Claude 3 Haiku & \$1.98 \\
    GPT-3.5 Turbo & \$2.73 \\
  \bottomrule
\end{tabular}
\end{table}

%%% CONCLUSIONS AND FUTURE WORK %%%
\section{Conclusions and Future Work}\label{sec:conclusions-future-work}

Our exploration into leveraging large language models for automating relevance judgment tasks in low-resource language scenarios, demonstrated using Tetun, has yielded results comparable to those achieved in high-resource languages, thus encouraging further research in low-resource languages (LRLs). The availability of freely and openly accessible models like LLaMA3 opens up possibilities for advancing relevance judgment tasks, particularly in low-resource language contexts, even with the limited digital content available on the web.

Our experiment demonstrated that despite LLaMA3's knowledge being limited to December 2023\footnote{\url{https://github.com/meta-llama/llama3/blob/main/MODEL_CARD.md}} and the availability of fewer than 45k Tetun documents on the web by that time \cite{kudugunta-et-al-2023, de-jesus-nunes-2024-labadain}, it achieved an agreement level comparable to high-resource languages like English. This indicates that automated relevance judgment tasks are feasible for other LRLs as well.

In future work, we plan to extend this research by incorporating a wider variety of examples in our prompts and testing with other freely and openly available models to compare the results. This approach will help validate and potentially expand the use of large language models in relevance judgment tasks.

%%% ACKNOWLEDGMENT %%%
\section{Acknowledgment}

This work is financed by National Funds through the Portuguese funding agency, FCT - Funda\c{c}\~{a}o para a Ci\^{e}ncia e a Tecnologia, within project LA/P/0063/2020 (DOI 10.54499/LA/P/0063/2020) and the Ph.D. scholarship grant number SFRH/BD/151437/2021 (DOI 10.54499/SFRH/BD/151437/2021).

%%% BIBLIOGRAPHY %%%
\bibliography{llm4eval}

%%% APPENDIX %%%
\appendix

\section{System Prompt Details}\label{annex:prompt}

Details of the system prompt used in the automatic relevance judgments, including four examples of query-document pairs along with the reasoning and the corresponding score for each.

\newtcolorbox[list inside=promptboxannex,auto counter,number within=section]{PromptBoxAnnex}{colback=gray!5!white, colframe=gray!75!black,title={Prompt \thetcbcounter: Details of the System Prompt.},label=promptboxannex}

\begin{PromptBoxAnnex}

You are an expert assessor and you are tasked with assessing the relevance between the input query and its corresponding document, assigning a score from 0 to 3. A score of 0 indicates irrelevant; 1, marginally relevant; 2, relevant; and 3, highly relevant.\\

Example 1:\\
\textbf{query:} ``Programa mestradu no pós-graduasaun UNTL''\\ \textbf{document:} ``Estudantes Pós-Graduasaun IOB Kuda Ai-Oan iha aldeia Payol no Bedois''\\ 
\textbf{reason:} ``The query is about postgraduate and master's courses at UNTL, whereas the document discusses the activities of postgraduate students from IOB. Although both query and document contain the term 'postgraduate', the query specifically is targeted courses at UNTL. Therefore, they are irrelevant.''\\
\textbf{score:} 0.\\

Example 2:\\
\textbf{query:} ``Kursu mestradu no pós-graduasaun UNTL''\\
\textbf{document:} ``Kursu Desportu UNTL sei realiza graduasaun dahuluk tinan ne’e''\\
\textbf{reason:} ``The query is about postgraduate and master's courses at UNTL, whereas the document focuses on a sports course. Despite both courses in the query and document being offered at UNTL, the sports course in the document is not specifically designed for postgraduate or master's levels. Thus, the document is only marginally relevant.''\\
\textbf{score:} 1.\\

Example 3:\\
\textbf{query:} ``Kursu mestradu no pós-graduasaun UNTL''\\
\textbf{document:} ``UNTL Nia Vise Reitór Asuntu Pós-Graduasaun No Peskiza Hakotu-iis''\\ 
\textbf{reason:} ``The document is relevant as it details the vice-director of the postgraduate program at UNTL. However, its relevance is somewhat diminished as it primarily discusses the unfortunate passing of the vice-director rather than the progress or implementation of the program. Hence, they are relevant.''\\
\textbf{score:} 2.\\

Example 4:\\
\textbf{query:} ``Kursu mestradu no pós-graduasaun UNTL''\\ 
\textbf{document:} ``UNTL Lansa Kursu Pós-Graduasaun No Mestradu Iha Área Lima''\\
\textbf{reason:} ``Both the query and document address postgraduate and master's courses at UNTL. The document strongly correlates with the query, containing the launching of postgraduate and master's courses at UNTL. Thefore they are highly relevant.''\\ 
\textbf{score}: 3.\\

The query and document to be evaluated are the following:\\\newline
\textbf{query:} $\{query\}$\\
\textbf{document:} $\{document\}$\\\newline
Your response must be in JSON format with the first field is ``reason'', explaining your reasoning, and the second field is ``score''.

\end{PromptBoxAnnex}

\end{document}